# Correlation of microwave surface impedance of $MgB_2$ thin film with material parameters and a temperature niche for microwave applications

**B. B. Jin[1], T. Dahm[2], C. Iniotakis[2], A.I. Gubin[3], Eun-Mi Choi[4], Hyun Jung Kim[4], Sung-IK Lee[4], W.N. Kang[5], S.F. Wang[6], Y.L. Zhou[6], A. V. Pogrebnyakov[7,8], J. M. Redwing [8], X.X. Xi[7,8], N. Klein [1]**

1) Institut für Schichten und Grenzflächen (ISG) and cni - Center of Nanoelectronic Systems for Information Technology, Forschungszentrum Jülich, 52425 Jülich, Germany

2) Universität Tübingen, Institut für Theoretische Physik, Auf der Morgenstelle 14, 72076 Tübingen, Germany

3) Usikov Institute of Radiophysics and Electronics, NAS of Ukraine, 12 Acad. Proskura Str., 61085 Kharkov, Ukraine

4) National Creative Research Initiative Center for Superconductivity, Department of Physics, Pohang University of Science and Technology, Pohang, 790-784, Korea

5) Department of Physics, Pukyong National University, Pusan, 608-737, Korea

6) Laboratory of Optical Physics, Institute of Physics, Chinese Academy of Sciences, Beijing, 100080, P. R. China

7) Department of Physics, The Pennsylvania State University, University Park, Pennsylvania 16802, USA

8) Department of Material Science and Engineering, The Pennsylvania State University, University Park, Pennsylvania 16802, USA

**Abstract**

Two issues related to the microwave surface impedance $Z_s$ of $MgB_2$ thin film are discussed in this Letter, both being significant for potential microwave applications. At first, a correlation between $Z_s$ and $\alpha = \xi_0/\ell$ was found, where $\xi_0$ is the coherence length, and $\ell$ is the mean free path. The surface resistance $R_s$ decreases with $\alpha$ at moderate and large values of $\alpha$ and saturates when $\alpha$ approaches one. The values of the penetration depth at zero temperature $\lambda(0)$ for different films could be well fitted by $\lambda_L (1+\alpha)^{1/2}$, yielding a London penetration

**depth $\lambda_L$ of 33.6 nm. The second issue is to find a temperature niche for possible microwave applications. Between 10K and 15K, our best MgB$_2$ films possess the lowest $R_s$ values compared with other superconductors such as NbN, Nb$_3$Sn and the high-temperature superconductor YBa$_2$Cu$_3$O$_{7-\delta}$.**

The superconductor MgB$_2$ has generated a great deal of interest because of its simple structure, its relatively high critical temperature ($T_c$) and a pronounced two-gap nature [1]-[4]. One of the potential applications of MgB$_2$ are high-frequency devices. To some extend, the relevance of MgB$_2$ can be elucidated by a systematic study of the microwave surface impedance [5]. In our previous work a sapphire resonator technique has been successfully employed to measure $Z_s=R_s+j\omega\mu_0\lambda$ of MgB$_2$ thin films at a frequency $f =\omega/2\pi= 18$ GHz [6][7]. Here $R_s$ is the microwave surface resistance, $\omega$ is the angular frequency, $\mu_0$ is the free-space permeability, and $\lambda$ is the penetration depth. In our present work we use the same method to investigate eight different MgB$_2$ thin films of different quality prepared by chemical vapor deposition (CVD) with post-deposition annealing [8], pulsed laser deposition (PLD) with in - situ annealing [9], and in situ hybrid physical-chemical vapor deposition (HPCVD) [10] methods. The three films deposited on MgO substrates by the CVD method are randomly oriented with a thickness of about 600 nm. The three films deposited on Al$_2$O$_3$ by PLD are c-axis oriented with a thickness of about 400nm. The remaining two films are epitaxial films deposited by HPCVD on Al$_2$O$_3$ and SiC, respectively, with thickness values of about 100 nm (on Al$_2$O$_3$) and 300 nm (on SiC), respectively. $T_c$ measurements show that all samples exhibit sharp transitions of the dc resistivity with a width of less than 0.3 K. For the sample on SiC, the highest $T_c$ of 41.3 K was observed.

From the measured temperature dependence of $Z_s$, the quantities $R_s(T)$, $\lambda(0)$, $\Delta_\pi(0)$ and $\rho_0$ were extracted according to a procedure described in Ref. [7] Here, $\Delta_\pi(0)$ is the value of π-band energy gap at zero temperature. $\rho_0$ represents the dc resistivity right above $T_c$ as being determined from the $R_s$ according to skin effect theory. Since the residual surface resistance $R_{res}$ ($R_{res} = R_s(T\rightarrow0)$) of the MgB$_2$ thin film is less than the resolution of $R_s$ for our setup (about 50μΩ [6]), the measured $R_s(T)$ represents just the change of $R_s$ with temperature.

Fig.1 shows the $\Delta_\pi(0)$ dependence on $\rho_0$. It should be noticed that $\Delta_\pi(0)$ roughly increases monotonically with $\rho_0$. Qualitatively, this observation is likely to reflect the two-gap nature of $MgB_2$: The increase of $\rho_0$ is accompanied by an increase of interband impurity scattering between the two bands, which results in a tendency to equalize the size of the two gaps. Thus, the small gap will increase with $\rho_0$. In the inset of Fig. 1 we show $\Delta_\pi(0)/T_{c0}$ as a function of $T_c/T_{c0}$. Here, $T_{c0}$ is the $T_c$ of the film on SiC, which is close to the clean limit.. The solid line shows the change of the small gap with $T_c$ based on interband impurity scattering theory [11] assuming $\Delta_\pi(0)/\Delta_\sigma(0) = 0.28$ ($\Delta_\sigma(0)$ is the value of the energy gap of the σ-band at zero temperature). Details of this analysis will be discussed in a later contribution.

The dependence of $Z_s$ on the material parameters $\lambda_L$, $v_F$, $\Delta_\pi(0)$ and $\ell$ or the two dimensionless quantities $\alpha = \xi_0/\ell$ and $\gamma(\ell=\infty) = \lambda_L/\xi_0$ can be described within BCS-theory[12]. Here, $\lambda_L$ is the London penetration depth, $v_F$ is the Fermi velocity, and $\ell$ is the mean free path. Because our study is restricted to MgB$_2$ samples only, $\gamma$ is a constant, and only the $Z_s$ dependence on $\alpha$ is presented. Here we only consider the π-band coherence length. There is another coherence length corresponding to the σ band. However, we have found in our previous work the conductivity is dominated by the π band [7]. According to BCS theory $\xi_0=\hbar v_F/\pi\Delta_\pi(0)$ [13], where $\hbar$ is Planck's constant, $v_F$ = 5.35×10$^5$m/s is the Fermi velocity of the electrons in the π band [14]. The mean free path can be calculated from $\rho_0$ by $l = v_F / \rho_0 \omega_p^2 \varepsilon_0$, where $\hbar\omega_p$ = 5.9 eV is the plasma energy [3] and $\varepsilon_0$ is free-space permittivity. From this analysis we found that the coherence length $\xi_0$ scatters between 43 and 53 nm, which agrees with the value of 49.6 nm determined by vortex imaging of single crystals [15]. Fig. 2 shows the $R_s$ dependence on $\alpha$ at $T$ = 15 K (solid) and 20 K (open). A monotonic increase can be found for large $\alpha$ values, which tends to saturate as $\alpha$ decreases. The 100 nm-thick epitaxial film on a sapphire substrate prepared by HPCVD deviates strongly from the general trend of the other samples. TEM analysis has demonstrated that there is a 30 - 40 nm thick non-superconducting layer between the $Al_2O_3$ substrate and $MgB_2$ thin film [10], which is expected to have a large influence on the properties of very thin films.

Fig.3 shows the measured $\lambda(0)$ as a function of $\alpha$. According to the BCS model, $\lambda(0)$ can be described by $\lambda_L(1+\alpha)^{1/2}$ [13]. The solid line represents the fit yielding $\lambda_L$ of 33.6 nm. The inset shows the $\lambda(0)$ dependence on $(1+\alpha)^{1/2}$, indicating a high level of confidence for the BCS model. The value of $\lambda_L$ determined by this analysis was found to be in excellent agreement with the plasma frequency used for band structure calculations [3]. This value also agrees roughly with the one obtained from microscopic calculations (=39.5 nm) in the clean limit[16].

In contrast to the observed $\lambda(0)$ dependence on $\alpha$, the observed $R_s$ dependence on $\alpha$ is still not fully understood. In particular, the effect of the anomalous coherence peak [7] requires further studies.

For conventional s-wave superconductors such as Nb ($T_c$ = 9.2K), NbN ($T_c$ = 16K) and $Nb_3Sn$ ($T_c$ = 18K) very low $R_s$ values can be achieved at working temperatures much lower than $T_c$, because $R_s$ is proportional to $\exp(-\Delta(0)/kT)$ for $T < T_c/2$. Above $T_c/2$, $R_s$ increases even more rapidly. On the other hand, the high-temperature superconductor YBCO has a finite $R_s$ value below a certain temperature and does not exhibit any exponential temperature dependence due to its d-wave nature [5]. Fig. 4 shows the temperature dependence of $R_s$ below 20 K for the samples with the lowest $R_s$ of each type of orientation. For comparison, the $R_s$ values of high-quality epitaxial YBCO films deposited on $LaAlO_3$ and r-cut sapphire, NbN and $Nb_3Sn$ films are also depicted [17][18][19]. One can easily see that $MgB_2$ has the lowest $R_s$ between 10 K and 15 K. From cryogenic cooling point of view, 10 to 15 K can make a big difference to 4 K, in particular with regard to cost and power efficiency of closed-cycle refrigerators.

Apart from very low $R_s$, the epitaxial film made by HPCVD on SiC has the highest $T_c$, the longest $\xi_0$ and the lowest $\lambda(0)$. These parameters are of great importance for electronic application. The high $T_c$ allows for high working temperatures, the large value of $\xi_0$ should make $MgB_2$ favorable with regard to the possible realization of tunnel junctions. On the other hand, the low values of $\lambda(0)$ indicate that a small layer thickness is sufficient for multilayer circuits and that kinetic inductance effects are rather small. Structural analysis has demonstrated that this particular film is composed of well-connected grains and has a smooth surface. Hence, this technique offers great promise for $MgB_2$ integrated circuits in the future with some potential to replace niobium.

In summary, our study clearly demonstrates a correlation between $Z_s$ and $\alpha$ in $MgB_2$ films, if material parameters describing the π-band are used. A temperature niche for microwave applications was found to exist between about 10 K and 15 K, where $MgB_2$ becomes competitive with other superconducting materials.

The authors thank Dr. R. Wördenweber for providing the $R_s$ data of YBCO thin film. One of the authors (A.I. Gubin) is funded in part within the INTAS program by the European Union, which supports his research stay in Jülich. The work at Postech is supported by the Ministry of Science and Technology of Korea through the Creative Research Initiative Program. The work at Penn State is supported in part by ONR under grant Nos. N00014-00-1-0294 (Xi) and N0014-01-1-0006 (Redwing).

**Figure Captions:**

Fig.1: The dependence of the π-band energy gap on resistivity of $MgB_2$ thin films determined from $Z_s$ measurements. The inset shows the π-band energy gap $\Delta_\pi(0)/T_{c0}$ as a function of $T_c/T_{c0}$. The solid line shows the dependence expected from interband impurity scattering in a two-gap superconductor.

Fig.2: The measured dependence of $R_s$ on $\alpha = \xi_0/\ell$ at 15 K (solid) and 20 K (open).

Fig.3: Experimental dependence of $\lambda(0)$ on $\alpha$ and a BCS fit yielding a London penetration depth $\lambda_L$ of 33.6 nm. The inset clearly shows the linear relation between $\lambda(0)$ and $(1+\alpha)^{1/2}$.

Fig.4: Comparison of surface resistance of $MgB_2$ with YBCO, NbN and $Nb_3Sn$ films. The $R_s$ curves of $MgB_2$ thin films represent the best $R_s$ values for films of different crystalline orientations.

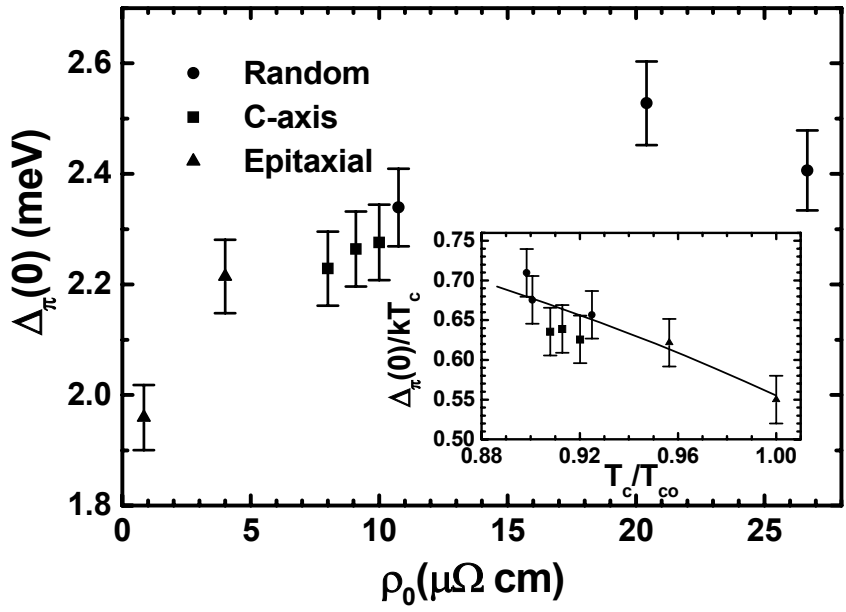

Fig.1

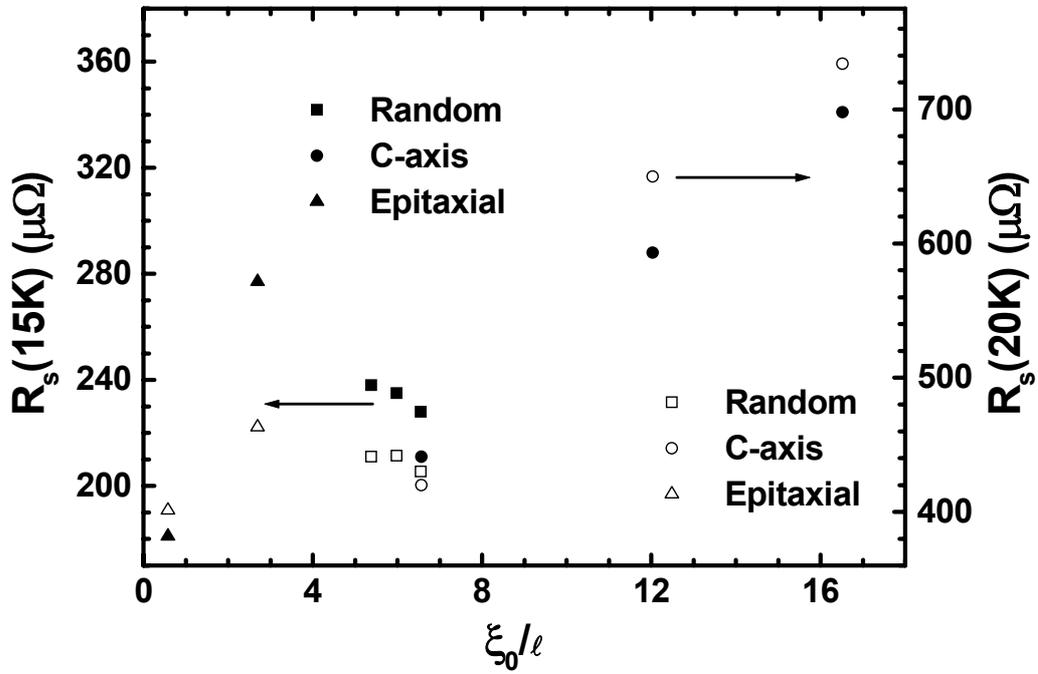

Fig.2

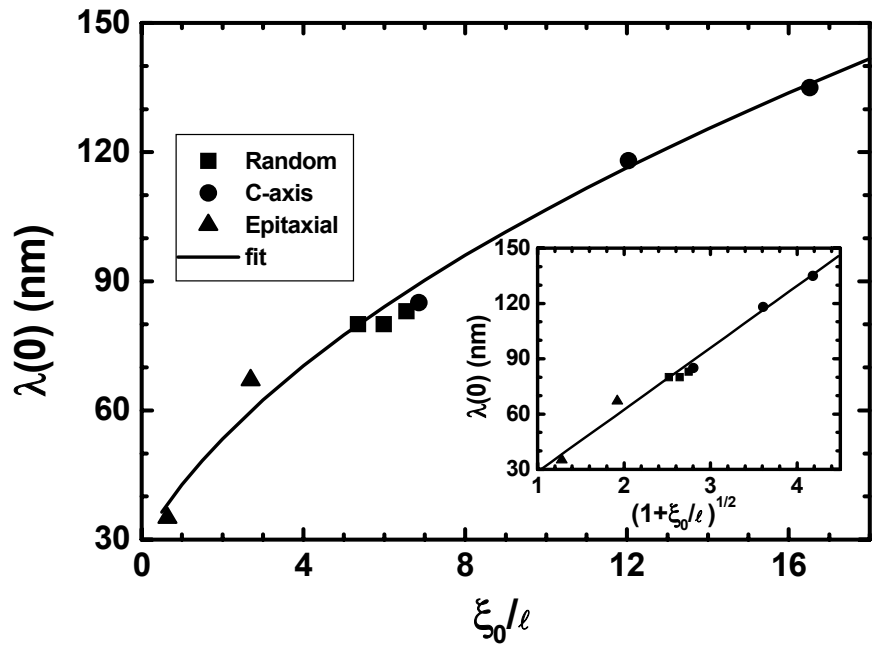

**Fig.3**

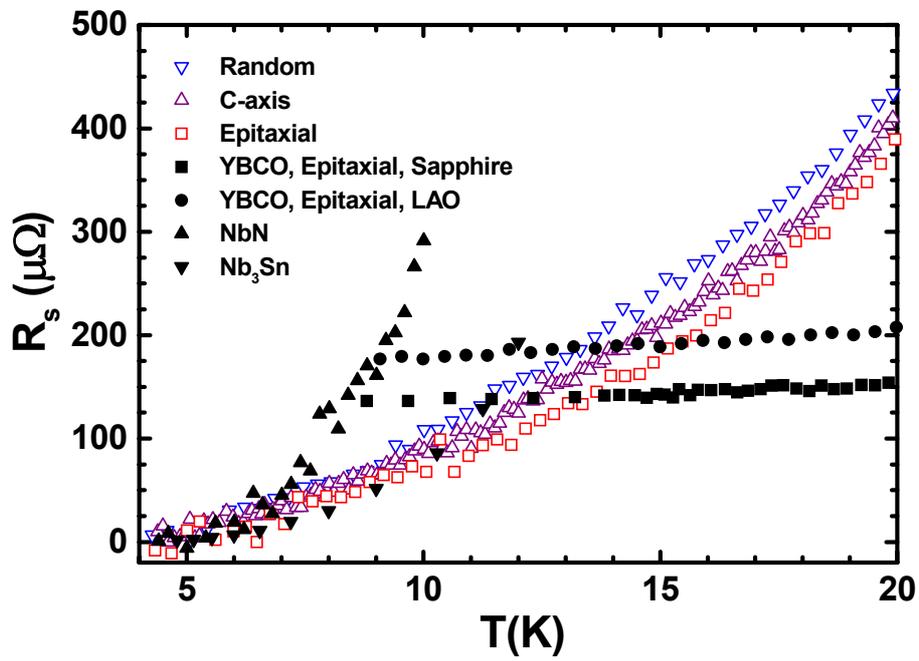

**Fig.4**